\documentclass[10pt,journal,compsoc]{IEEEtran}
\ifCLASSOPTIONcompsoc
  \usepackage[nocompress]{cite}
\else
  \usepackage{cite}
\fi
\usepackage[cmex10]{amsmath}
\usepackage{siunitx}
\usepackage{cite}
\usepackage{psfrag}
\usepackage[utf8]{inputenc}
\usepackage[T1]{fontenc}
\usepackage{amsmath,amsfonts,amsbsy,amssymb}
\usepackage{mathabx}
\usepackage{mathrsfs}
\usepackage[nolist]{acronym}
\usepackage{tabularx}
\usepackage{amssymb}
\usepackage{amsmath}
\usepackage{graphicx}
\usepackage{cite}
\usepackage{multirow}
\usepackage{wasysym}
\usepackage{multirow}
\usepackage{float}
\usepackage{xcolor}
\usepackage[font=footnotesize]{caption}
\usepackage{algorithm}
\usepackage{algorithmic}
\usepackage{footnote}
\usepackage{array}
\usepackage{rotating}
\usepackage{nomencl}
\makenomenclature

\begin{document}
\author{
Amin Hekmatmanesh,
Pedro H. J. Nardelli,~\IEEEmembership{Senior Member,~IEEE,}
Heikki Handroos\IEEEcompsocitemizethanks{\IEEEcompsocthanksitem The authors are with School of Energy Systems, LUT University, Finland. This paper is partly supported by Academy of Finland via: (a) ee-IoT project n.319009, (b) FIREMAN consortium CHIST-ERA/n.326270, and (c) EnergyNet Research Fellowship n.321265/n.328869.}
}
%\address[1]{}

\title{Review of the State-of-the-art on Bio-signal-based Brain-controlled Vehicles}

\IEEEtitleabstractindextext{
\begin{abstract}
Brain-controlled vehicle (BCV) is an already established technology usually designed for disabled patients.
This review focuses on the most relevant topics on the brain controlling vehicles, especially considering terrestrial BCV (e.g., mobile car, car simulators, real car, graphical and gaming cars) and aerial BCV, also named BCAV (e.g., real quadcopter, drone, fixed wings, graphical helicopter and aircraft) controlled using bio-signals such as electroencephalogram (EEG), electrooculogram and electromyogram.
For instance, EEG-based algorithms detect patterns from motor imaginary cortex area of the brain for intention detection, patterns like event related desynchronization\textbackslash event related synchronization, state visually evoked potentials, P300, and generated local evoked potential patterns.
We have identified that the reported best performing approaches employ machine learning and artificial intelligence optimization methods, namely support vector machine, neural network, linear discriminant analysis, k-nearest neighbor, k-means, water drop optimization and chaotic tug of war optimization optimization.
We considered the following metrics to analyze the efficiency of the different methods: type and combination of bio-signals, time response, and accuracy values with the statistical analysis.
The present work provides an extensive literature review of the key findings of previous ten years, indicating the future perspectives in the field.
\end{abstract}

}
\maketitle

%\mbox{}
\nomenclature{EEG}{Electroencephalogram}
\nomenclature{EOG}{Electrooculogram}
\nomenclature{EMG}{Electromyogram}
\nomenclature{LDA}{Linear Discriminant Analysis}
\nomenclature{PSD}{Power spectrum Density}
\nomenclature{BVC}{Brain-Vehicle Control}
\nomenclature{BCA}{Brain-Controlled Aerial Vehicle}
\nomenclature{BCI}{Brain Computer Interface}
\nomenclature{GHMM}{Gaussian mixture-hidden Markov model}
\nomenclature{SVM}{Support Vector Machine}
\nomenclature{SMSVM}{Soft Margin SVM}
\nomenclature{RBF}{Radial Basis Function}
\nomenclature{GRBF}{Generalized RBF}
\nomenclature{fNIRS}{functional Near-Infrared Spectroscopy}
\nomenclature{fMRI}{functional magnetic resonance imaging}
\nomenclature{ERD}{Event Related Desynchronization}
\nomenclature{ERS}{Event Related Synchronization}
\nomenclature{ERP}{Event Related Potentials}
\nomenclature{SSVEP}{state visually evoked potentials}
\nomenclature{ICA}{Independent Component Analysis}
\nomenclature{PCA}{Principal Component Analysis}
\nomenclature{LEP}{Localized Evoked potential}
\nomenclature{DFA}{ Detrended Fluctuation Analysis}
\nomenclature{WDO}{Water Drop Optimization}
\nomenclature{IM}{Imaginary Movement}
\nomenclature{RP}{Readiness Potentials}
\nomenclature{SNR}{Signal to Noise Rate}
\nomenclature{SNR}{Global Positioning System}
\nomenclature{CSP}{Common Spatial Pattern}
\nomenclature{HbO}{Oxy-hemoglobin }
\nomenclature{HbR}{Deoxy-hemoglobin }
\nomenclature{CTWO}{Chaotic Tug of War Optimization}
\nomenclature{RLDA}{Regularized Linear Discriminant Analysis}
\nomenclature{NN}{Neural Network}
\nomenclature{K-NN}{K-Nearest Neighbor}
\nomenclature{GGAP}{Generalized Growing and Pruning}
\nomenclature{RBFNN}{Radial Basis Function Neural Network}
\nomenclature{QN}{Queuing Network}
\nomenclature{CNN}{Convolutions Neural Network}
\nomenclature{MPC}{Model Predictive Control}
\nomenclature{LR}{Logistic Regression}
\nomenclature{GMM}{Gaussian Mixture Model}
\nomenclature{HMM}{Hidden Markov Model}
\nomenclature{TP}{True Positive}
\nomenclature{TN}{True Negative}
\nomenclature{FP}{False Positive}
\nomenclature{FN}{False Negative}
\nomenclature{EBC}{Emergency Brakes Control}
\nomenclature{OAC}{Obstacle Avoidance Control}
\nomenclature{EP}{Evoke Potential}
\printnomenclature

\section{Introduction}\label{sec:introduction}
\label{sec:introduction}

The recent research in neuroscience supported by the development of high-precision sensors and artificial intelligence methods has significantly increased our knowledge about how the brain works.
In particular, human body movements activate the neurons in the sensorimotor cortex area. The activated neurons generate action potentials for different actions in which has different patterns with specific properties. Several studies have had performed for exploring the patterns in the Electroencephalogram (EEG) signals. Thereafter, automatic methods of identifying and predicting the patterns specifically onset of a voluntary movement have been started \cite{von1867handbuch}.

\begin{figure*}[t]

  \begin{center}
  \includegraphics[width=1.6\columnwidth]{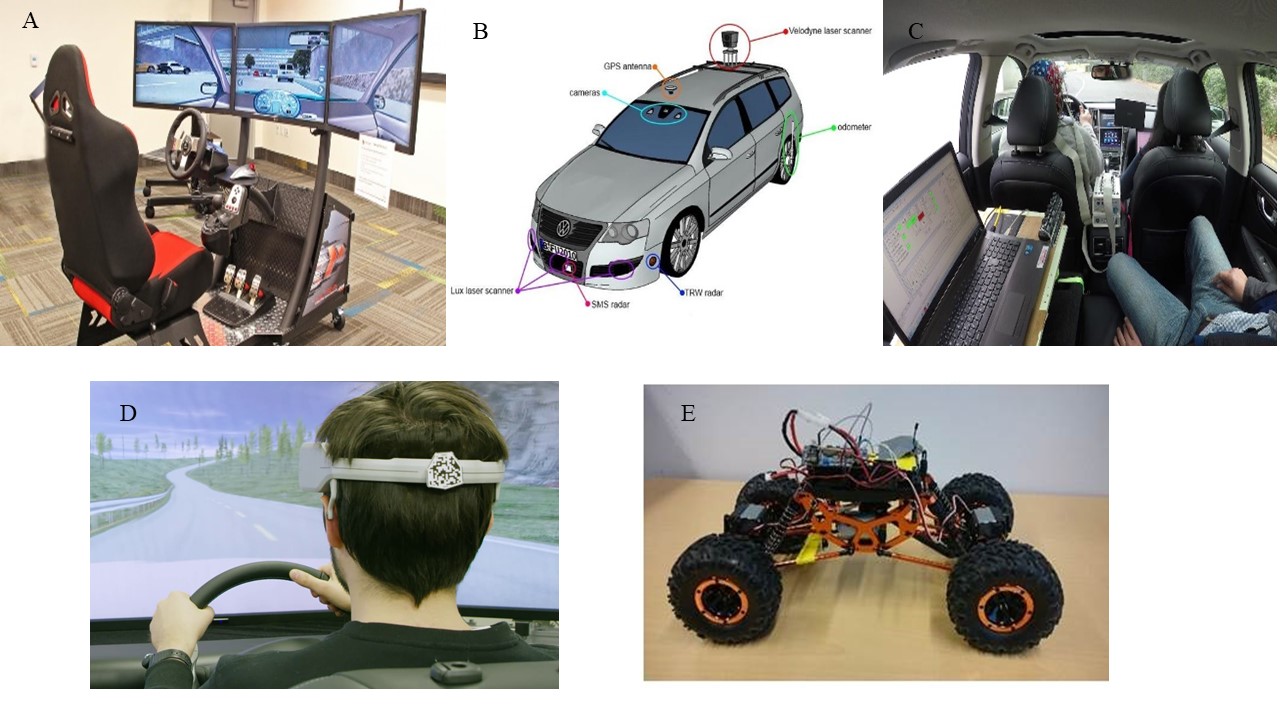}
  \caption{Different BCV applications: A. is a vehicle simulator, B. is a vehicle with different external sensors and camera, C. is a real vehicle, D. is a video game vehicle, E. is a mobile vehicle controlled by EEG}
  \label{fig:BCVexamples}
  \end{center}
  \vspace{-0.5cm}
\end{figure*}

The Brain Computer Interface (BCI) science use the patterns for controlling applications such as bionic hand, bionic leg, mobile robots, vehicles and typing. Among BCI's vast variety of applications, this review focuses on the Brain-Controlled Vehicle (BCV) and Brain-Controlled Aerial Vehicle (BCAV) applications mainly designed for normal people and specifically brain stroke disabled patients. One of the most important bio-signals is the EEG and the first step is knowing the EEG rhythms and changes after tasks and stimulation. The important patterns for diagnosing the intention of the drivers are Event Related Potentials (ERPs), State Visually Evoked Potentials (SSVEP),  Desynchronization\textbackslash Event Related Synchronization (ERD\textbackslash ERS), Readiness Potentials (RP) and Local Evoked Potentials (LEP).
Figure \ref{fig:BCVexamples} illustrates the BCV applications.

EEG is a real-time signal that the current resolution is not good enough  for the BVC and BCAV applications. Therefore, hybrid methods have been developed to cover the deficits of previous methods. For example, using EEG with other bio-signals such as Electromyogram (EMG), Electrooculogram (EOG) and functional Near-Infrared Spectroscopy (fNIRS) to gain more information from human for controlling applications. In addition to bio-signals, external sensors are employed for recording and analysing the environment information for better analysing the EEG and the situation.

In specific, BVC aims at tasks related to car navigation, namely keep the lane, passing the cars, following cars, turning, Obstacle Avoidance Control (OAC), braking in different situations, specifically the Emergency Brakes Control (EBC). The same commands are computed for the BCAV with two more directions of moving upward (take off) and downward (landing).
\hfill

By a real intention of movements, specific patterns appear in the EEG about 0.5s to 2s earlier and then the intention turn to action \cite{crammond2000prior}. The concept of the reviewed studies are developing novel algorithms for finding the onset of Imaginary Movement (IM) patterns such as ERD\textbackslash ERS and Readiness Potentials (RP).
\hfill

The aim of the present review is preparing a comprehensive review on the BVC and BCAV studies in the previous 10 years. Also, we expect that the present contribution would be helpful for understanding the recent history of the field, how the ideas and studies have been developed and improved. Then, new ideas for the future developments, which are based on the recent technologies, could be better contextualized.
The papers we have covered in this study are summarized in Table \ref{tabel1} to Table \ref{tabel4}, presented in Appendix, to provide a systematic comparison between the different contributions. 

The rest of this paper is divided as follows:
Section \ref{sec:background} we provide the background knowledge, mainly based on brain rhythms with intention identification approaches and  data acquisition model, both applied in BCV and BCAV, as well as limitations and open questions.
Section \ref{sec:driver-intention} discuss the 
algorithms for automatically predicting the drivers intention based on patterns from bio-signals.
Section \ref{sec:studies} presents applications for training and testing the models in real-time mode.
Section \ref{sec:concl} concludes this review by posing the already solved questions and current limitations, while providing our future vision about the topic.

\section{Background on BVC}
\label{sec:background}
\subsection{Brain Rhythms and Patterns for BVC and BVCA}

Brain is a multi-functional system that different neurons generate different rhythms with specific features. The detectable rhythms changes based on the the type of actions, stimulation and task experiments. Changes of the rhythms is also a key clue for early diagnosing of a disease and  serious situations. By focusing on the sensorimotor cortex area rhythms it is possible to predict the subject’s intention of movement. Some of the studied patterns for intention detection (thinking) are ERD \cite{neuper2006event}, ERP \cite{sokhadze2017event}, ERS \cite{neuper2006event}, and SSVEP \cite{kemp2002steady} in which they are defined as follows.

\subsubsection{ERD\textbackslash ERS Pattern}
ERD is a cognitive pattern, in which appears after intending to move and ERS is the second pattern appear immediately after the ERD if the intention turns to action. The location of recording the pattern is the sensorimotor cortex area of the brain \cite{hekmatmanesh2019combination}.
%, Figure \ref{erd}.

\subsubsection{SSVEP Pattern}
The SSVEP is a response pattern, in which appears when a visual stimulation applied on a human. By applying a visual stimulation in a specific range the same evoked potential patterns named as SSVEP will appear in the visual cortex. The advantages of the SSVEP is the high Signal to Noise Rate (SNR) in compare to other patterns. \cite{kemp2002steady}.

\subsubsection{ERP Pattern}
ERPs are the measured electrophysiological response by EEG to a specific stimulation. The P300 is a known brain response to a cognitive event  after 300 ms. Some of the other patterns are N100, N200 and p100 etc. The P300 is the aim  pattern in the controlling applications \cite{neuper2006event}.
%, Figure \ref{p300}.
As an example, the P300 pattern has been used for typing (predicting, decision making) applications for disabled patients, by concentrating on the letters. In BCV applications, the P300 is employed for destination selection.

\subsubsection{LEP Pattern}
Some studies searching for new ERPs for better controlling system. Therefore, new tasks are designed and applied to stimulate neurons other than sensorimotor cortex area such as auditory tasks, then the obtained patterns are employed for further computations and controlling applications\cite{coenen2015uav}.

\subsubsection{RP Patterns}
RP is a pattern generated before real movement about 1.5s to 1s. The RP pattern is a useful pattern for repetitive voluntary movements such as walking. In the processing, RP is divided into early and late RPs. The early appears about 1.5s before voluntary movement at the central area of the cortex and the late RP appears about 500ms before the voluntary movement at the primary motor cortex area \cite{kornhuber1965hirnpotentialanderungen,brunia2003cnv}. 

\subsection{Data Acquisition}
In order to control a BCI application using the bio-signals, amplifiers for measuring the human body changes during the experiments is required. The well-known devices are EEG, EMG, EOG amplifies (are suitable for real-time processing) and fNIRS and functional Magnetic Resonance Imaging (fMRI) devices that the details are presented as follows:

\subsubsection{EEG, EMG, EOG Amplifiers}
In order to measure noninvasive signals from heart, brain activities and muscles ECG, EEG and EMG amplifiers are manufactured, respectively. The usual electrodes for acquiring EEG, EMG and ECG signals are the Ag\textbackslash AgCl that known as nonpolarized electrodes. The other popular electrode is disposable electrode, named as gel-based or Bio-Potential (BP) electrode which is one time use. In theory, the BP electrode senses ion flow on the tissue surface, and then converts it to electron current. For the EMG measurement using the BP electrodes, the ion distribution generates through applying the nervous stimuli and muscles contraction.
%
%\subsubsection{Electrodes Quality}
The employed electrodes categorized as nonpolarized and polarized. The nonpolarized electrodes (Ag\textbackslash Agcl) pass the current across the electrolyte interface. Therefore, less noise records in compare to the polarized electrodes in case of movement noise. Also, the nonpolarized electrodes are easy for manufacturing and it has very low half-cell potential named as dc offset. Therefore, the Ag\textbackslash Agcl is popular for the EEG recording in compare to other electrodes.
The polarized electrodes do not let the current moves freely across the interface between the electrode and the electrolyte in which acts similar to capacitors.  

\subsubsection{fNIRS}
fNIRS is a noninvaseivean imaging system for measuring the hemoglobin (Hb) concentration changes in neuro-vascular of the brain. The Hb concentration changes is measured by optical intensity measurements (characteristic absorption spectra) through the near-infrared light. The studies employed fNIRS are usually hybrid methods with EEG signals for real-time controlling of the BVCA applications. The fNIRS has been used for the primary motor cortex area for imaginary tasks to find the precise executed areas and using them for identification procedures \cite{naseer2015fnirs}. 

\subsubsection{fMRI}
The fMIR is an accurate noninvaseive imaging system for demonstrating the localized power in a brain map with high resolution. The mechanism is working based on the hemodynamic changes of the brain which is associated with neuronal activity \cite{abreu2018eeg}. In the present review, the fMRI is employed for controlling BCAV applications. The fMRI is usually employed as a hybrid method with EEG to obtain high results in real-time systems. 

\subsubsection{External Sensors}
Hybrid methods are a combination of different signals to increase the efficiency of the results. In some methods, a combination of different bio-signals with non-bio-signal are employed for identifying the driver's intention and navigating accurately such as combination of the EEG with EMG, Global Positioning System (GPS), camera, fNIRS, google glasses and motion sensors named as external sensors (acceleration ,velocity, wind speed, etc.)\cite{gohring2013semi,bi2018novel}.

\subsection{Open questions, limitations and research topics}
In order to control a vehicle, either BCV or BCAV, using brain signals following main steps are required: %
\begin{enumerate}
    \item pre-processing,
    \item feature extraction,
    \item optimization
    \item feature selection,
    \item classifiers,
    \item statistical analysis,
    \item real-time experiment.
\end{enumerate}

\begin{figure*}[t]
  \begin{center}
  \includegraphics[width=1\textwidth]{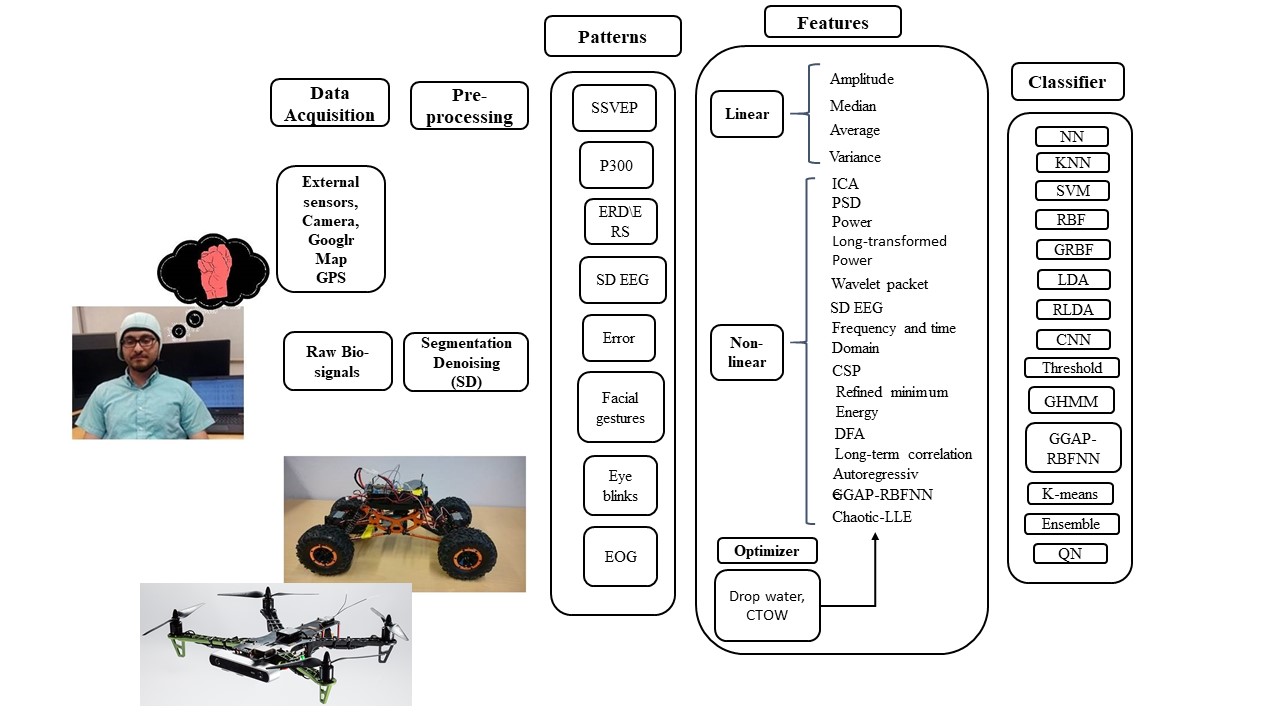}
  \caption{The applied methods (features and classifiers) in an algorithm to identify the drivers intention for BCV and BVAC applications. A sample of controlling a mobile vehicle and drone application for training and testing in illustrated. }
  \label{fig:general}
  \end{center}
\end{figure*}

Figure \ref{fig:general} provides a more detailed description of those steps, and possible options to be considered that will be described next.

The improvements of the mathematical identification methods in the challenges of controlling vehicles through bio-signals and the road map of improvements during the years 2010 to 2020 are considered. 
The initial BCV EEG-based questions and limitations were how to find the source of the neuron’s activation, frequency range of neurons activities, the specific patterns related to the applied stimulation and develop algorithms for finding the patterns automatically. Since now, many of them are solved.

The unsolved problems are mathematical algorithms for noise rejection and identifying specified patterns automatically with high precision. More specifically, the most challenging section is developing effective algorithms for feature extraction and classification for automatic pattern identification. Further questions raised of which neurons are connected in a specific task and how neurons communicate after the stimulation, the topic, named as neuron’s connectivity. 

The next ongoing challenges are the mathematical approaches for predicting patterns, designing real-time algorithms and speeding up processing of the time-consuming methods such as wavelet-based methods.
The key problems in the BCV applications based on the EEG are (i) the nonlinearity of the brain, in which generate patterns with different varieties for individual participants; (ii) the denoising the affected EEG signals by white noise (which is highly nonlinear and includes whole frequencies and it is similar to EEG); (iii) hardware limitations (distance and speed) communication for portable and wireless devices (irrespective of Bluetooth and Wi-Fi) in the real-time applications. 

In the following, we explain the methods applied to identify the intentions of drivers based on brain signals.

\section{Driver's Intention Identification}
\label{sec:driver-intention}

To detect and predict the driver's intention for controlling a BCV and BCAV, the steps presented in the previous subsection needs to be followed. 
We will provide a brief review of each of them throughout this section (supported by the compilation presented from Table \ref{tabel1} to Table \ref{tabel4} at Appendix.

\subsection{Preprocessing}
The preprocessing is an important step to remove unwanted signals named as noise from data and prepare it for further processing. The preprocessing consist of filtering and segmentation. Different studies based on their aims various filters are employed. For example, EEG is filtered to obtain the Alpha waves, \cite{bi2013using} used a filter bank to extract main frequency waves such as Delta band (1-4 Hz), Theta band (4-8 Hz), Alpha band (8-14 Hz), Beta band (14-30 Hz), and low Gamma band (30-60 Hz) \cite{teng2017eeg}, and filtered EEG for frequency range of 8-16 Hz to extract the ERD\textbackslash ERS patterns \cite{hekmatmanesh2020combination,hekmatmanesh2019combination}. 
In the next step, the denoised signals in the aim frequency ranges are employed for feature extraction. 

\subsection{Feature Extraction}
Feature extraction algorithms are one of the most important steps of the identification algorithms. A good feature is a feature that shows high distinction for a specific part of a signal against other part of the signal. Features are categorized as linear and nonlinear. Some of employed features for BCV and BCAV are average, median, power, amplitude, variance, PSD, FFT, autoregressive, long term correlation, cross correlation, spectral amplitude, frequency filtered signal such as alpha wave, Common Spatial Pattern (CSP), Independent Component Analysis (ICA), FastICA, wavelet, Detrended Fluctuation Analysis (DFA), chaotic algorithms such as largest lyapunov exponent, HbO and HbR (hemoglobin concentration changes) for fNIRS etc. In some algorithms, the initial values of the features required optimization.

\subsubsection{Water Drop Optimization (WDO)}
\label{DWO}
WDO is an evolutionary developed algorithm which is based on the water river behavior to find it’s way. The WDO's aim is searching for optimum values in functions using the water behavior in the river. The idea of the algorithm is constructed based on two water characteristics during moving, 1- velocity and 2- number of conveyed soils with water. The advantage of this approach is high speed converging \cite{shah2009intelligent}. This algorithm is also useful in optimization classification approaches.

\subsubsection{Chaotic Tug of War Optimization (CTWO)}
\label{CTWO}
 CTWO is a recent developed optimization algorithm with the concept of competition in rope pulling. The CTWO select two teams as solution candidates for applying in pulling forces (interaction between teams), the amount of forces is relative to the quality of solutions. In the algorithm, it has five steps of 1- initialization 2- weight assignment 3- competition 4- new generation 5- termination. The advantage of the CTWO is higher speed in comparison with the stochastic searches \cite{kaveh2017optimum,hekmatmanesh2019eeg}. This algorithm is also effective for optimizing classifiers.

In some algorithms, the extracted features are needed to be selected by feature selection approaches.

\subsection{Feature Selection}
In some cases of data recording, noise affect the quality of data highly. Therefore, the extracted features combined with noise, which are needed to be selected. The employed feature selection algorithms for BCV and BCAV are Principal component Analysis (PCA) and Linear Discriminant Analysis (LDA). The selected features are then fed to the classifiers for categorising.

\subsection{Classifiers}
\label{classifiers}
In this section, the developed methods as classifiers are considered. The utilized classifiers were used in three modes of offline, semi-autonomous and real-time modes. Offline mode is useful for training a model for the real-time mode, but in some researches only the efficiency of the methods in offline mode is presented. Real-time mode is the aim of BCI studies to find out how much the developed algorithms are reliable. The semi-autonomous applications is usually combination of real-time mode and automatic methods based on external sensors which are not under control of users. In another words, only some limited commands are under control of user and the rest are automatic.\hfill

Classifiers are divided into three types of supervised, semi-supervised and unsupervised. In the presented reviews studies, the employed algorithms for the BCV and BCAV applications are supervised and unsupervised. In the supervised algorithms, the labels of segmented signals for different sates are determined, but in unsupervised classifiers the labels are enigmatic. In the following steps, the employed classifiers for controlling a BCV and BCAV are presented and the efficiency of individuals are presented in Table \ref{tabel1} to Table \ref{tabel4}.

\subsubsection{ K-Nearest Neighbor (K-NN)}
K-NN is a supervised multi classifier, which is based on voting. The voting computations is based on the Euclidean distance of the features around (neighbours) a feature. The number of neighbors is the K parameter which is very important. The results are highly related to the number of selected K neighbors and the right value is defined usually by test and test. For example, by selecting K=7, the algorithm select the seven closest data to the new coming data. If at least four of the neighbours belongs to Class 1 then the algorithm vote the new data to be classified as Class 1, otherwise categorized in class 2.

\subsubsection{LDA and Regularized LDA (RLDA)}
LDA is a linear classifier, which is also used for feature selection by reducing the feature space dimension. The concept of the algorithm is maximizing between-group scattering over within-group scattering. In another words, the algorithm searches for the projections that reduce the inter class variance whilst increase the distance between classes. The decision maker is a hyperplane, which is applied on the mean of the two classes distance \cite{fukunaga2013introduction,hekmatmanesh2020combination,shakeel2015review}.\hfill

The fundamental concept of the RLDA is the same as LDA that the regularization mode enable ability of employing more correlated features at the same time with less error. The concept of the regularizing the LDA algorithm is regularizing the scattering of the inter class features to have a non-singular matrix. The LDA is then applied on the regularized matrix. The RLDA is also fast processing and generate efficient results.

\subsubsection{Neural Network (NN)}
NN is a nonlinear supervised classifiers which is suitable for multi classifications. The structure of NN with back propagation is based on the assigned weights for the neurons that are nonlinear functions such as Sigmoidal and Gaussian. The structure is so flexible to be changed according to the features distribution and results.

\subsubsection{RBFNN and GGAP-RBFNN}
RBFNN is a supervised classifier, which is based on the RBF activation function. In the algorithm, the RBFNN contains input-, hidden- and output- layers with the activation functions (neurons). The neurons trained with the labeled features in a training procedure. For classifying a new coming data, the neurons compute the difference between the new coming data and trained weights. The new data classified as the same as most similar neuron which is belongs to.\hfill

The RBF is flexible to be used as a kernel of classifiers such as SVM (defined in the next part). A generalized version of the RBF named as Generalized RBF (GRBF) is introduced and employed as a kernel in SVM. The GRBF is parametrized by three free parameters of width and center in Gaussian function. The results are more stable and higher accuracies obtained, when it is used as a kernel \cite{noori2014k,hekmatmanesh2014sleep}.\hfill

In order to increase the efficiency, speed and optimize the network size of the RBFNN a method named as generalized growing and pruning (GGAP) is developed. The GGAP algorithm, link the aim desired accuracy of the RBFNN with the importance measurements of the closest added new neuron. The importance measurement computed using average content of the specific neuron \cite{zhao2019identification}.

\subsubsection{Support Vector Machine(SVM) and Soft Margine SVM (SMSVM)}
SVM is a supervised binary classifier. In the procedure, the the input data is transferred into higher dimension by kernels such as RBF, GRBF, polynomial etc and then a linear decision maker applied. In cases of multi classification a technique of one versus all is employed. For example, in a case of three classes, first class 1 is classified against features of class 2 and 3 as one class, and then class 2 is classified against class 3.\hfill

Different versions of SVM is also employed in the classification studies such as SMSVM, the base on the same as SVM but an optimization algorithm is set to customize the cost function and regularization algorithm. The SMSVM is more efficient algorithm with the GRBF kernel \cite{hekmatmanesh2020combination,hekmatmanesh2019combination}.

\subsubsection{Threshold}
Thresholds play an important role in the threshold-based classifiers. Employing only thresholds as classifiers has risk of high error rates in the results, specifically for the nonlinear features. In order to consider the risks of the threshold-based classifiers algorithms such as Vector Phase Analysis (VPA) are employed. VPA is a threshold-based classifier, which is employed for hemoglobin concentration changes (HbO and HbR) in the fNIRS features.

\subsubsection{Queuing Network (QN)}
QN is an analytic method for solving equations efficiently. The QN developing idea was constructing models for predicting waiting time in queues. In a BCV study, the QN predictive model is employed for controlling steering in a vehicle \cite{bi2016queuing}. The QN for BCV is constructed by preview, predict and control modules. The input of the preview module is the desired path to determine the desired vehicle position. The predictive module input is the road information and the vehicle states, which are obtained by external sensors to determine the predictive position. The control module input is the subtraction of the preview from the predictive module to compute the error for the steering command computations.  

\subsubsection{Convolutions Neural Network (CNN)}
CNN is a supervised classifier, which is based on deep learning and widely employed in image processing. The CNN algorithm is divided into three parts of convolution- and pooling- layers, and feed forward NN. A constructed CNN has ability of having several convolution and pooling layers that the number of layers needs to be adjusted. The convolution layer is employed for producing features from the input data. The pooling layer is then employed for dimension reduction of the convolution layer and the NN is for classifying.

\subsubsection{Ensemble Classifier (EC)}
EC classifier is a combination of several learning algorithms in a classifier. It is then known as a generalized new approach to increase the efficiency of any classifiers in comparison with individual classifiers such as boosting and bagging methods. Boosting is a popular method in EC for reducing bias to obtain strong dependency in the data. In the study \cite{zhuang2019ensemble}, a multi-class boosting method is employed, named as AdaBoost \cite{freund1995desicion}.  

\subsubsection{Model Predictive Control (MPC)}
MPC is a method for controlling a process meanwhile satisfying equation criterion. The remarkable advantages of the MPC are flexibility and open formula for linear, nonlinear and multi variable equations without changing the MPC controlling algorithm. In a BCV application, the MPC employed for predicting the driver’s intention based on a model \cite{lu2019model}.

\subsubsection{Logistic Regression (LR)}
The LR is a variable dependent binary linear supervised classifier, which is based on a logistic function (S-shaped function) in a statistical model. The LR model’s aim is modeling the probability of the features related to individual classes such as imagination of right and left hand movements. In another words, the LR mission is finding a linear decision boundary between classes using the parameters which are assigned to the features. The computations are based on relation between dependent binary variables of the classes and maximum likelihood estimation. The weights are then adjusted and applied on the features for classification.

\subsubsection{K-means}
K-means is an unsupervised clustering algorithm. In the algorithm, K number of clusters are defined and then the mathematical approaches find the K centers which are center of clusters. Each data is then clustered based on the distance to the centers. The clusters are formed by minimizing the within-cluster variance.

\subsubsection{Gaussian Mixture Model (GMM)}
GMM is an unsupervised clustering method for categorising features into C number of classes. In the algorithm, it assumes that the features of each class has a Gaussian distribution and the feature space that consist of several classes follows a rule of mixing finite Gaussian distributions and each Gaussian has specific center and width. Therefore, GMM is a  probabilistic model in which features are distributed based on a mixture of a number of Gaussian functions with unknown parameters that are computed by Maximum Likelihood Estimation.

\subsubsection{Hidden Markov Model (HMM)}
The HMM concept comes form the interpretation of the world problems that comes from visible and invisible sources. The HMM is an unsupervised clustering, which is the extension of the Markov Model (MM), that the principals are based on the Markov Chain (MC). In another definition, HMM is a statistical model, which is based on observable patterns that are relative to unobservable interior factors. In the procedure of the HMM, the observable patterns and unobservable internal factors are named as patterns and states, respectively. The algorithm has two random processes for the layers, named as hidden and visible processes for the hidden states and observable patterns, respectively. The hidden states compute a MC and the probability distribution of the patterns relative to the states. The features will be then categorized Based on the probability computations \cite{yoon2009hidden}.

\subsubsection{Performance Computations}
In order to measure the efficiency of the classifiers for comparison, the statistical algorithms such as accuracy, sensitivity and specificity are employed that the computations are based on on four parameters as follows: True Positive (TP), True Negative (TN), False Positive (FP) and False Negative (FN). Sensitivity is the TP rate and Sensitivity is the TN rate in classifications. TP is the correct features which are categorized correctly, TN is the the correct features which are categorized incorrectly, FP is the false features (incorrect) which are categorised correctly and FN is the false features which are categorized incorrectly \cite{noori2014k,hekmatmanesh2017sleep,hekmatmanesh2014sleep}.

\subsection{Studies on Real-Time Mode Applications}
In different studies disparate applications are employed. Even if the rule of controlling are approximately the same, but each application employed tricks to obtain better accuracy. The real-time BCV and BCAV employed applications in the reviewed papers are presented as follows: vehicle simulator, graphical game, real car in real world, mobile robot, quadcopter, drone, helicopter and aircraft.
In the following section, we will describe in details studies about controlling BCV and BCAV.

\section{Studies on BCV and BCAV}
\label{sec:studies}
In order to control a vehicle different methods using the bio-signals have been employed. Published studies on the BCV and BCAV topics are driver’s intention detection for controlling vehicles for navigation, changing lane, steering control, \cite{lu2019combined,hekmatmanesh2019optimized}, EBC \cite{hernandez2018eeg,zhuang2019ensemble}, and OAC \cite{bi2018novel,teng2017eeg}. Some studies reported the accuracy results based on individual subjects that we compute the average values and report them in Table \ref{tabel1} to Table \ref{tabel4}. 

\subsection{Employed Techniques and BCV Efficiencies}
In a key series study, Haufe et al. \cite{haufe2011eeg} implemented an EBC system for BCV applications using EEG and EMG signals in a graphical racing car task. In the algorithm, the ERP patterns relative to the emergency brakes were employed for train the algorithm and then tested on a simulated vehicle. In the algorithm, the computed features were the area under the curve and then classified using the RLDA classifier. The evaluations for braking in real-time mode were based on the detection accuracy and response time (reaction) parameters. Then, in their next identification study, Kim et al. \cite{kim2014detection} implement an algorithm for an assistant brake (soft and sharp braking) for different situations specifically EBC based on the driver’s intention. The experiment was performed using a simulated vehicle which is combined with the EEG and EMG devices. The braking scenario for the simulated vehicle is designed in a graphical traffic street. The idea in the algorithm was extracting features from three patterns as follows: combination of RPs (time interval from 300 ms before stimulation to 600 ms after stimulation), MI (ERD\textbackslash ERS- obtained by filtering EEG data between 5 and 35 Hz) and ERP (obtained by Hilbert transformation) patterns. Results showed that the detection results increased in comparison with the previous study. Also, the authors noticed that neurons generate specific ERP patterns relative to emergency cases. Therefore, Haufe et al. \cite{haufe2014electrophysiology} identified the ERP patterns related to the emergency brakes in a task, and then improved the robustness of the previous methods in a real-time experiment \cite{kim2014detection, haufe2011eeg}. The aim of the study was employing the designed algorithm in \cite{kim2014detection} in a real-world experiment based on different tasks. In Haufe et al. study, the experimental tasks were auditory and vehicle-following for predicting the emergency brakes based on ERPs in a non-traffic area. The algorithm and trained classifier is then employed for a real-world (real-time) experiment. However, the accuracy and robustness values did not report.\hfill

In a different study by Gohring et al. \cite{gohring2013semi}, designed an experiment for controlling a semi-autonomous vehicle. In the experiment, a set of 16 external sensors with a camera is combined for navigating the vehicle on a road semi-automatically, except braking and steering commands. For controlling the vehicle, steering and braking, EEG signal is recorded in two different scenarios for OAC and driving. The commands for controlling the steering and braking states are computed using the ERD\textbackslash ERS patterns. The utilized camera and external sensors were significantly helpful for decreasing the Evoke Potential (EP) detection error rates.  The algorithm is then applied on a real vehicle that improved the results, but the complete mathematical methods did not present.\hfill

In a continuous study of different groups for controlling a vehicle, different tasks were designed and EEG patterns were employed. Bi et al. \cite{bi2013using} designed a head-up display task based on the SSVEP pattern for controlling a vehicle simulator. In the experiment, the first step was identifying the alpha waves using the LDA classifier to turn the vehicle on and off. Next, a vehicle navigation (turn right, left and move forward) based on SVM classifier is employed for the OAC algorithm for an offline mode. The trained algorithm is then tested in a real-time mode. Results for the OAC and turn the vehicle on and off were significant, but the results for the three directions navigation showed high variant accuracies. Limitations for this study were small number of participants, the response time did not consider and the recommended speed for this algorithm is 30-40 Km. As a continuous study, Bi et al \cite{bi2013head} employed the P300 pattern for selecting the driver’s intention destination for the same simulated vehicle in \cite{bi2013using}. Results showed higher accuracy with double number of participants in comparison with the previous study. In continuous, Fan et al. \cite{fan2014brain} combined their previous methods and application using the SSVEP pattern and alpha EEG waves for controlling a vehicle simulator with the following commands: starting, stopping, staying in lane, OAC and curve control. In the algorithm, the PSD features were computed from the patterns, and then categorized using the LDA algorithm. Then, Bi et al. \cite{bi2016queuing} proposed a model for controlling the BCV steering on the same application in \cite{bi2013using} and \cite{bi2013head}. In the algorithm, a model designed, named as queuing network for considering the performance of driver’s intention identification (using the SSVEP pattern) in advance. The identified commands employed for predicting the driver’s intention for going forward, turning left and right. The designed algorithm was based on velocity, acceleration, road information and vehicle position in a road. The performance of the model reported based on the accuracy that is increased in compare with their previous above-mentioned studies. Regarding the results, the response time for the model reported 500 ms and accuracy results were not stable because of high variation between subjects. In the next study, Bi et al. \cite{bi2018novel} developed an identification method for emergency brakes. In the experiment, a set of external sensors in an embedded system was employed to analyze the environment condition. The implemented algorithm in comparison with the previous methods in \cite{bi2013using,bi2013head,bi2016queuing} for emergency brakes was more complicated and the obtained accuracy and time response results increased significantly.\hfill

Some algorithms are effective and need adjusting by optimization algorithms. Teng et al. \cite{teng2017eeg} implemented an algorithm for EBC in an OAC simulated task. In the algorithm, a model consists of feature extraction and classification designed utilizing a combination of two approaches: CSP features and RLDA classifier. The novelty of the algorithm is using an optimization method for feature selection, named as the sequential forward floating search. The algorithm also tested in pseudo-online mode and results showed significant ascending accuracy rate and descending response time. Then, Bi et al. \cite{bi2018novel} improved Teng et al. method by increasing safety of the BCV through external sensors for analyzing the environment information and detecting obstacles. Results showed increasing response time increased, but accuracies remained as the \cite{teng2017eeg} study. Regarding one recent successful study, Bi et al. \cite{bi2018novel} employed a combination of external sensors and road information for autonomous driving to increase driver’s safety. Bi et al. \cite{bi2018novel} study has potential of increasing driver’s safety significantly. As a future world-wide road map, to expand the Bi et al. method, the road model needs to be connected to a database such as google map to receive the road information with high speed connection, a millisecond delay will cause a disaster. Therefore, utilizing external sensors has ability of updating the road database for improving the model accuracy and updating the road information. Regarding the study considerations, the next telecommunication generations has potential of solving the distance and speed constraints impressively.\hfill

A continues series of studies has been implemented to evaluate the driver's emotions when they are nervous and relax. Zhang et al. \cite{zhang2015eeg} designed a real-time algorithm based on error-related potentials to control a simulated and real vehicle. In the experiment, authors employed the PSD features based on the filtered error-related ERP signals. The LDA classifier is then applied for controlling speed, lane change and dynamic of a vehicle. Results were not impressive in compare to the last studies. Next, Yang et al. \cite{yang2018driving} designed an algorithm for predicting the driver’s behavior (aggressive and unaggressive behavior) using methods of lateral- (changing lane) and longitude control (speed acceleration) as a driver assistant model. In the algorithm, the computed features were amplitude, long-transformed power, PSD from different frequency bands. The method's novelty was designing two recognition layer in the algorithm that consists of two supervised (SVM, KNN) and one non-supervised learning (K-means) classifiers. Then, Zhuang et al. \cite{zhuang2019ensemble} implemented an EEG-based algorithm with online visual feedback for controlling a simulated vehicle for a BCV application. Zhuang et al. employed the combination of wavelet and Canonical correlation analysis (CCA) for filtering data, and then two methods, named as the ensemble model, CSP with the SVM and CNN classifiers were employed to identify the MI patterns. The task was controlling a vehicle in three states of right- and left- steering and acceleration for the OAC task. Although, the wavelet is known as a time-consuming algorithm and causes delay, authors did not mention the delay in the real-time experiment.\hfill

In a recent continuous study, optimization algorithms were also employed for adjusting chaotic features with various classifiers for controlling a mobile vehicle. Hekmatmanesh et al. \cite{hekmatmanesh2018common, hekmatmanesh2020combination} implemented a method for controlling a mobile vehicle for moving forward, braking states and the same method has been applied on a prosthetic hand. In the procedure, several classifiers with the Filter Bank CSP (FBCSP) and Discrimination Sensitive Learning Vector Quantization (DSLVQ) training algorithm were employed for considering the effectiveness of the DSLVQ coefficients and finding the best classifier for nonlinear features. Results showed that the SMSVM classifier using the generalized RBF (GRBF) kernel obtained the best results in comparison with the traditional SVM, KNN and NN classifiers. Then, the best classifier employed in their further studies. The GRBF kernel were used in their previous bio-signal processing studies \cite{hekmatmanesh2014sleep, noori2014k}. The limitation of \cite{hekmatmanesh2020combination} was employing the CSP, which is useful for only two classes and the extended CSP approaches for multi classes increases the error rates significantly. In another study, Hekmatmanesh et al. \cite{hekmatmanesh2019combination, hekmatmanesh2019investigation, hekmatmanesh2019optimized} employed nonlinear features for detecting the ERD\textbackslash ERS patterns. In \cite{hekmatmanesh2019combination, hekmatmanesh2019optimized}, a remote vehicle controlled using the XBEE Bluetooth connection chipset. In the algorithm, customized mother wavelets were designed based on the ERD\textbackslash ERS patterns of individual subjects, and then imported to the wavelet packet algorithm automatically. Then, the wavelet packet integrated with the detrended fluctuation analysis (DFA) method to compute the long-term correlation. The features were classified using the best selected classifier in \cite{hekmatmanesh2020combination}. The results for controlling the vehicle for braking and moving forward improved in comparison with predefined mother wavelet methods. Wavelet is a time-consuming method which is the main limitation. The computed delay in a real-time system, obtained between one to two seconds. On the other hand, the wireless hardware limitation was distance connectivity, which was about 12 meters. In the next step, Hekmatmanesh et al. \cite{hekmatmanesh2019eeg, hekmatmanesh2017sleep} attended to improve nonlinear features in chaotic theory for controlling a BVC based on the ERD\textbackslash ERS patterns. In the algorithm, the Largest Lyapunov Exponent (LLE) was computed, and then the initial values were optimized using the WDO \cite{hekmatmanesh2019optimizing} and CTWO \cite{hekmatmanesh2019eeg} optimization intelligent methods. Results improved in comparison with the normal LLE only in offline mode.\hfill

In a recent key series studies, for increasing the SNR and accuracy recognition rate several training algorithms are combined in two layers. Lu et al. \cite{lu2018eeg} designed an algorithm based on longitude control system for controlling speed of a simulated vehicle. In the algorithm, CSP is employed for augmenting the EEG signal SNR, and then PSD features were extracted (from the SSVEP patterns) and classified using the SVM classifier with RBF kernel. Accuracy results showed high accuracy variation for individual subjects. Later, Lu et al. \cite{lu2019combined} extended the longitude control by combining the lateral control for the simulated vehicle through the EEG. In another word, the idea is extending two classes to four classes with the same identification classifier. The driver’s intention task was changing lane, selecting path and following cars. Accuracy results showed high variation for subjects. Next, Lu et al. \cite{lu2019model} developed a method named as MPC to increase the performance. The algorithm was designed based on penalty values, which are computed using the cost function for safety criterion. The MPC controlling model was introduced by the same authors in \cite{bi2016shared}. The algorithm was combination of two virtual scene scenarios, controlling road-keeping test and OAC. The reported achieved results showed high variations for the algorithm performance.\hfill

Combining supervised and unsupervised classifiers for obtaining a better results is also a valuable idea. Zhao et al. \cite{zhao2019identification}, attended to design models for driver’s intention in case of braking. Therefore, a combination of the Gaussian Mixture-Hidden Markov Model (GHMM) and GGAP with RBFNN (GGAPRBFNN) is employed for increasing accuracy and decreasing time response. The algorithm used for identifying slight- and normal- braking states and then tested in a real vehicle. The obtained results compared with their previous work \cite{wang2018study} that was significant, but the time response did not consider.\hfill

In order to have a complete review, Stawiki et al. \cite{stawicki2016driving} designed a method for the BCV systems based on the SSVEP patterns. In the experiment, it is attended to control a mobile vehicle with a graphical user interface and camera for live feedback system. The novelty of the algorithm was utilizing a computational approach to remove noise and increase the amplitude of the SSVEP patterns before feature extraction. The achieved results for the large number of participants were significant. Hernandez et al. \cite{hernandez2018eeg}, designed an algorithm for identify vehicle brake in different conditions based on the EEG signals. In the experiment, a vehicle simulator with OAC and emergency braking scenario is prepared. Then, time domain features with the SVM and CNN classifiers were categorized, the time domain features were the preprocessed EEG signals. The obtained response time for braking in emergency cases (high speed) were not significant. The obtained results did not show significant accuracy changes in comparison with the other studies with the same aim. In another recent study, Nguyen et al. \cite{nguyen2019detection} developed a method for identifying driver’s intention for EBC in a vehicle. In the experiment, a simulated vehicle is utilized. The algorithm consists of the EEG band power and auto-regressive features with a NN classifier. Results obtained showed significant accuracy and response time improvement. \hfill

The second interesting vehicle is the aerial vehicles. Many attempts have been performed to control BCAV using bio-signals. In the next part, the employed methods for controlling a BCAV is considered. 

\subsection{Employed Techniques and BCAV Efficiencies}
\begin{figure*}[t]
  \begin{center}
  \includegraphics[width=0.75\textwidth]{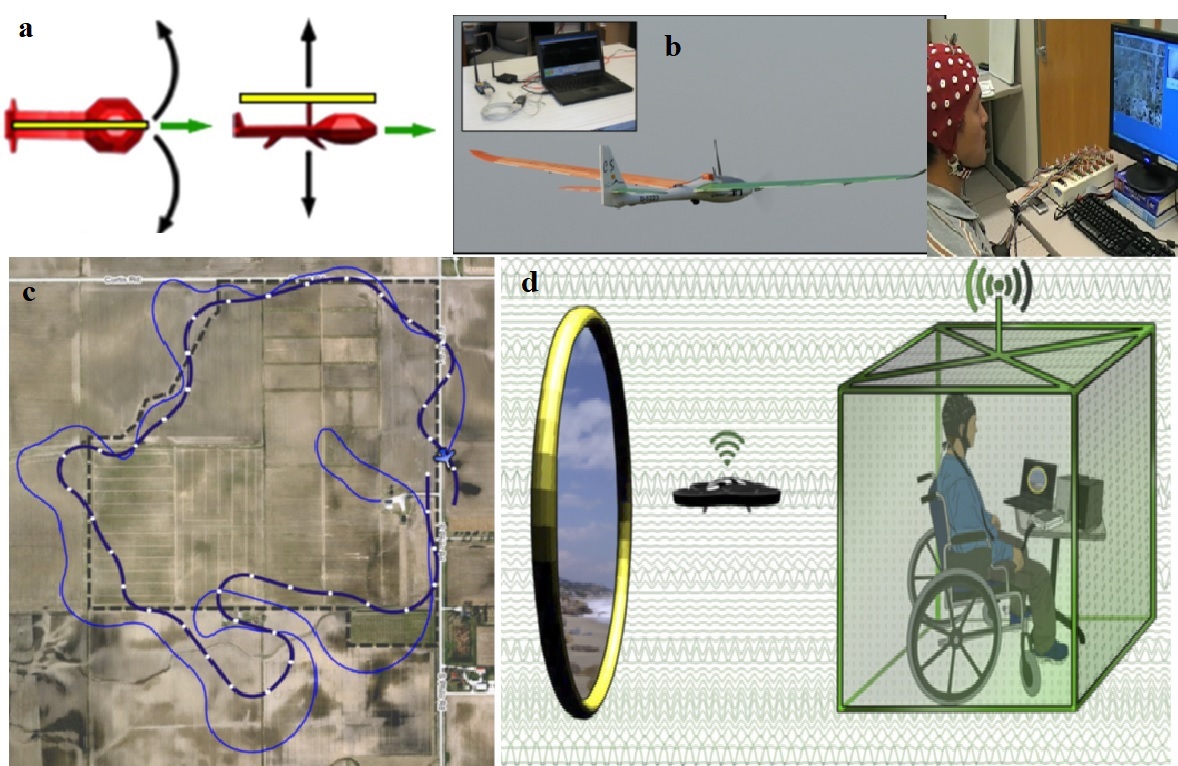}
  \caption{Some employed applications for BCAV, a. virtual helicopter, b. fixed wings c. controlling based on following a selected path, d. drone.}
  \label{BCAV}
  \end{center}
\end{figure*}

% aerial vehicles
In the present review, the second type of vehicles named as aerial vehicles such as drones, quadcopters, helicopters and airplanes. BCAV applications are the other interesting topics for investigation, Figure \ref{BCAV}. Recently, a review of aerial vehicles in 2018 \cite{nourmohammadi2018survey} is published, in which focused on studies before 2015 and studies between 2013 to 2015 are mostly conferences. In review \cite{nourmohammadi2018survey}, the focus is on categorizing types of aerial vehicles and controlling methods, in which employed in our considerations. Here, we present a complementary methodological review based on bio-signal processing on the effective studies from 2010 to 2020.\hfill

Recently, drone technology is a commercialized application and many industries and organizations have been employed to increase their productivity and efficiency. The combination of the unmanned aerial vehicles and BCI is a newborn idea, which is the goal of researches to gain the benefits of it. Advantages of using drones are significant, for example low cost producing, low cost transferring, low cost maintenance, ready to fly quickly, very productive for the conditions that pilots cannot fly, clean energy, difficult applications such as spacecraft applications \cite{allison2012towards}, etc. There are many categories of the aerial vehicles for different missions such as health care and military. In the present survey, the focus is reviewing on bio-signal processing techniques for controlling BCAV applications in non-military applications. For considering the other types of the aerial unmanned vehicles and structures please refer to \cite{allison2012towards}.\hfill

Recent, generation of hybrid methods have been employed for controlling aerial vehicles such as EEG, (f)MRI and (f)NIRS measurements. The fNIRS and the fMRI methods has limitation of real-time mode usage, but they have high resolution. On the other hand, the EEG has ability of utilizing in the real-time mode, but it does not have the same resolution as in the fNIRS and fMRI. Therefore, some studies combine the advantages of both techniques at the same time and introduce hybrid methods such as EEG with the fNIRS \cite{zafar2018drone, khan2017hybrid}, EEG with fMRI and EEG with the eye tracker \cite{kim2014quadcopter} etc. In the EEG-based aerial controlling algorithms, the following patterns are employed for feature extraction: ERD\textbackslash ERS, ERPs, SSVEP, eye movements and blinking. The computed features for the above-mentioned patterns are cross correlation, LR, mean, peaks and PSD which are classified with different classifiers such as SVM \cite{zafar2018initial, zafar2018drone} and LDA \cite{khan2017hybrid}.\hfill

The difference between BCV and BCAV is four navigation commands such as take off (up for drons), landing (down for drons) rotates for drones (differ from turning). The preliminary commands for navigation of a fixed wings and helicopters was controlling four main directions after manually take off. In a continuous studies, Royer et al. \cite{royer2010eeg} attended to control a graphical helicopter in four main directions using the ERD\textbackslash ERS patterns. In the algorithm, the extracted features were cross correlation and difference of the autoregressive spectral amplitude between right and left hemisphere. Then, features were categorized using a linear classifier. Two weak points of the algorithm were delay of 2.1 sec reaction time and low precision. In another study, Akce et al. \cite{akce2010remote}, used the ERD\textbackslash ERS patterns for controlling a fixed-wing aerial vehicle. In the experiment, a controlling algorithm based on selecting trajectory of a path through a binary classifier was designed. The obtained results were not impressive. Doud et al. \cite{doud2011continuous} improved Royer et al. \cite{royer2010eeg} study results, using control of a virtual helicopter for six directions based on time-frequency analysis and PSD features. Finally, the helicopter accuracy results improved by Lafleur et al. \cite{lafleur2013quadcopter}. In Lafleur et al. study, the helicopter was controlled in six directions based on the MI patters. The evaluations were based on a modification of information transfer rate approach.\hfill
% The quadcopter connected with WiFi and virtual feedback, which is provided by a camera in the quadcopter.

In a continuous hybrid studies, Kim et al. \cite{kim2014quadcopter} attended to control a quadcopter in eight directions based on nine points eye gazing training using eye tracker and EEG signals. In the algorithm, the camera data is first employed to extract eye's pupil features and then power of the EEG data and EOG paradigms are computed. The feature are then classified using the SVM classifier with a linear kernel. As a different application, Shi et al. \cite{shi2015brain} controlled a hex-copter using the ERD\textbackslash ERS patterns, cross correlation features and LR classification. In the application, a live camera is employed for the OAC to obtain better results. Later in an impressive study, Coenen et al. \cite{coenen2015uav} used different technique of response to mental task patterns for controlling a drone in two directions. The signal was recorded in an auditory imagination and spatial navigation mental task. The generated different patterns were key of improving the results. Next, kosmyna et al. \cite{kosmyna2015towards,kosmyna2015adding} attended to control a quadcopter for three directions using a hybrid EEG and EMG bio-signals. In the task left hand, right hand and foot tapping were employed to generate the patterns for controlling of turn right, left and move down. In the algorithm, the MI patterns with the facial patterns were extracted from the EEG and EMG signals, respectively. The features were then identified using the KNN algorithm and adaptive recurrent NN classifiers.\hfill

Some studies employing the same BCV experimental tasks for producing different patterns. kryger et al. \cite{kryger2017flight} controlled an aircraft simulator for six direction using EEG. In the experiment, only one subject is participated. Unfortunately, the authors did not mention the algorithms and mathematical methods in the study. In another study, Wang et al. \cite{wang2018wearable} proposed a method based on the four different flickering LEDs for controlling a quadcopter in four directions. The authors employed the Head head-mounted device (HMD) in the virtual task. In order to identify the SSVEP patterns the canonical correlation analysis (CCA) with a threshold classifier is employed. The reported results based on only threshold classifier is interesting.\hfill

Recently, hybrid method of EEG and fNIRS become a successful approach for controlling a BCAV application. Recording the EEG and fNIRS signals in real-time mode is new proposed method to increase the precision of the results. The first series of hybrid by EEG, fNIRS studies performed by Lin et al. \cite{lin2015implementing} for controlling a quadcopter for six directions. In the algorithm, EEG and EMG signals were recorded in a facial gesture task. The features were computed based on the EEG signals and then numbers of features were reduced using the PCA algorithm. Results were not reported as a significant achievement. Another hybrid study is performed by Zhang et al. \cite{jacksonquadcopter} to improve Khan et al. \cite{khan2015hybrid} idea. For controlling the quadcopter in six directions, a Google glass with the EEG signals was utilized. In the experiment, a task based on head posture MI is designed, then the ERD\textbackslash ERS and SSVEP were extracted for feature computations. The spectral features are then computed and then selected by PCA. The features were then classified using the SVM classifier with the RBF kernel. The identification algorithm was also the combination of 14 external sensors for the navigation. The results were not reported as significant achievements.\hfill

The second series of hybrid studies (EEG with fNIRS) which are combined with BCV techniques is performed as follows: Khan et al. \cite{khan2015hybrid} utilized the EEG and NIRS data for controlling a quadcopter. In the task, data was recorded based on the MI for two directions and a combination of the ERD\textbackslash ERS and SSVEP patterns was employed for feature extraction. In the algorithm, the PSD features were extracted from the SSVEP patterns and the oxygenated and deoxygenated  hemoglobin features were extracted from the NIRS data. The obtained results based on the hybrid data recording were significant. Next, Khan et al. \cite{khan2016hybrid} used the EEG, EOG and FNIRS data for extracting features for four directions and tested in a real-time experiment. In the algorithm, IM of left hand, left- and right- eye movements were employed for navigating the quadcopter with live video feedback. Also, an OAC algorithm was then developed using the SSVEP patterns. The method is applied on three subjects and then improved in \cite{khan2017hybrid}. The developments of Khan. et al. \cite{khan2017hybrid} were generating new patterns based on different stimulation such as Mental arithmetic, mental counting, word formation, and mental rotation stimulation, and increase the number of features peak, skewness, mean and power from the EEG, mean, peak, slope, peak, minimum and skewness from the fNIRS. The results were increased in compare to their previous method in \cite{khan2015hybrid}. Next, Khan et al. \cite{khan2017hybrid} improved the EEG-fNIRS method by exceeding the number of decoded commands to eight commands (clockwise and unclockwise rotations added) for controlling a quadcopter. In the experiment, two LDAs were employed for classifying. One LDA were used forclassifiying the fNIRS features and the other LDA used for the EEG features. Results showed significant success for classifying the eight quadcopter states. Later the same team in another \cite{zafar2018drone}, employed only the fNIRS signals to control the quadcopter for one state of moving forward. In the algorithm, mean, slope, peak, changes of HbO and HbR, HbT and COE were employed as features and then categorized by SVM, threshold circle and vector phase analysis classifiers. The weak point is the limited algorithm for two classes identification with 2.3s delay. Therefore, Zafar et al. \cite{zafar2018initial} performed another study for controlling the drone for three states of up, down and move forward. Therefore, mental  arithmetic  and  mental counting tasks were employed with the same classifiers and features in \cite{zafar2018drone}. An impressive average accuracy result were achieved but the delay was remained 2.3s. Recently, Kavichai et al. \cite{kavichai2019shared} improved the time delay in \cite{khan2017hybrid} using Shared Control Strategy (SCS) method. The SCS method employed environment information using external sensors. Therefore, Kavichai et al. combined the fNIRS and EEG features with three following measurements: eye movement, distance (measurement sensor), and Global Positioning System (GPS). Finally, through the fNIRS four commands and through the EEG four other commands were computed. The aim of the approach was OAC and reducing time response delay.\hfill

However, the challenge points, for BCV and BCAV named as finding fixed pattern (continuous varying EEG patterns of drivers causes variation in accuracy), response time, class identification precision, and robustness are still existing problems and causes faults in the system.\hfill

\section{Final Discussions}
\label{sec:concl}

In the presented BCI studies, bio-signal patterns have been employed for controlling BCV and BCAV applications. One of the main parameters in the BCV- and BCAV-based applications is reducing the delay of response time for processing and sending commands. Therefore, solving the delay of processing in real-time systems with high accuracy is an important limitation that needs more study. The second critical issue is the limitation of distance for communication systems in the BCV and BCAV applications.\hfill

New technologies supported by beyond 5G systems (e.g., \cite{moioli2020neurosciences})  have great potential for higher quality of communication with virtually no delay as in real-time processing. The high speed communications enable the applications to load high amount of data in the cloud/edge servers to store and use them within strict time constraints. Also, it is more applicable to use the road information through internet for different applications.
By developing the BCV and BCAV methods further, the issue of reliability of the application's security \cite{singandhupe2018reliable} will become a crucial point for future research. 

%\subsection{Conclusion}
All in all, detecting driver’s intention for emergency braking is a challenging point in the real world while stress, fatigue, mental workload, different emotions and environmental noise exists and varies individually. The second challenging in the vehicle emergency braking projects based on the bio-signals is the response time. The question is how much time is needed to prevent a crush in different speeds that needs more considerations. Despite the above-mentioned limitations, identifying the emergency brake situation based on the EEG has high risks. Emergency cases such as obstacle avoidance has limitations: 1- identifying an obstacle is different from predicting an obstacle, which is a difference between real dangerous situation or something with potential of danger, 2- environment has high negative influence on the results that increase the risk rate.

\bibliographystyle{IEEEtran}
\bibliography{ref2.bib}

\appendix

We present in this appendix a systematic presentation of the most significant literature in the topic of BCV and BCAV from the previous 10 years. 

	\begin{table*} 
  	    \centering
	    \caption{First part of reviewed studies in the past 10 years}\label{tabel1}
	    \begin{turn} {90}
		\begin{tabular}{|c|c|c|c|c|c|c|c|c|}
			\hline 
			 \shortstack{Authors\\Year} & Signal & Patterns & Features & Classifiers & \shortstack{Task and control\\commands}& Average Accuracy & RT & Modes
			 \\
			\hline
			\shortstack{Haufe et al. \cite{haufe2011eeg}\\2011} & EEG\&EMG & ERP & \shortstack{1- symmetric negative\\deflection in occipito-\\temporal area,\\2- negativity at\\central scalp sites\\3- positive deflection\\around electrode\\ CPz} & RLDA & \shortstack{combination of EEG,\\EMG with Gas\&EBC} & \shortstack{83.00\% passing\\collision} & 130ms & OM\&POM
			\\
			\hline
			\shortstack{Gohring et al. \cite{gohring2013semi}\\2013} & \shortstack{EEG,\\16 external\\sensors\\camera} & ERD\textbackslash ERS & \shortstack{pre-processsed\\EEG signal}& Threshold & \shortstack{R,L,S,B for\\Path selection,\\braking\&steering} & 90.00\%& \shortstack{26\% less than\\2 sec and 36\% \\between 5-10s} & OM, SAM 
			\\ 
			\hline 
			\shortstack{Bi et al. \cite{bi2013using}\\2013} & EEG & \shortstack{SSVEP,\\alpha waves} & PSD & LDA\&SVM & \shortstack{L, R, B, S for start,\\stop, stay in lane,\\OAC and curve\\control} & \shortstack{SSVEP detection=\\76.87\%\\alpha wave detection=\\93.53\% }& 2 sec to 3.5 sec & RTM, OM\\ 
			\hline 
			\shortstack{Bi et al.\cite{bi2013head}\\2013}& EEG& P300& PSD& LDA\&SVM & \shortstack{L,R,S for\\destination\\ selection}& 93.60\%$\pm${1.6\%} & 12 sec& RTM,OM 
			\\
			\hline 
			\shortstack{Kim et al. \cite{kim2014detection}\\2014} &EEG\&EMG & \shortstack{RP, ERD\textbackslash ERS\\ERP (visual evoked\\potentials\&p300)}&\shortstack{EEG from 300ms\\before stimilation\\to 600ms after stimul-\\ation for RP\\pattern, filtered\\ EEG between 5Hz\\ and 35Hz for\\ERD\textbackslash ERS and Hilbert\\ transfor-mation\\for ERP feature} & RLDA& \shortstack{RFM, SIM\&ES\\for sharp and soft\\braking intention} &-& 150ms & OM
			\\ 
			\hline 
			\shortstack{Haufe et al. \cite{haufe2014electrophysiology}\\2014}& EEG\&EMG & ERP& \shortstack{The same as \cite{haufe2011eeg}} & RLDA& \shortstack{the same as \cite{haufe2011eeg}}&-& \shortstack{RTM=200ms,\\RVM=237ms} &\shortstack{RVM on a\\non-public\\test}
			\\
			\hline 
			\shortstack{Fan et al. \cite{fan2014brain}\\2014}& EEG &P300\&SSVEP& \shortstack{0–512ms EEG\\time interval\\after P300\\onset}& LDA & \shortstack{destination\\selection}& \shortstack{OM = 99.07$\pm${0.40},\\RVM = 98.93\%}& \shortstack{OM RT\\selection =\\24.19s$\pm${0.85s},\\RVM RT=\\25.95s$\pm${1.04s}\\,no response time}& \shortstack{OM,RTM\& \\RTM}
			\\ 
			\hline 
			\shortstack{Zhang et al. \cite{zhang2015eeg}\\2015} & EEG\&EOG& \shortstack{filtered error\\-related potential}& PSD & LDA &\shortstack{L,R\&S for\\controlling speed,\\lane change and\\dynamic of vehicle} & \shortstack{OM=69.80\%$\pm${6.50\%}\\RVM=68.20\%$\pm${5.90\%}}&-& \shortstack{OM,RTM\&\\RVM}
			\\
			\hline
			\shortstack{Bi et al. \cite{bi2016queuing}\\2016}&\shortstack{EEG and road\\information}& SSVEP& \shortstack{Frequency-domain\\features}& \shortstack{Extended\\queuing network}& \shortstack{L,R\&S for turning\\left and right and\\going forward} & 88.77\%$\pm${13.5\%}& \shortstack{500 ms for model\\processing, RT\\not mentioned} & OM\&RTM
			\\
			\hline			
			\shortstack{Stawiki et al. \cite{stawicki2016driving} \\2016}& EEG& SSVEP & \shortstack{refined minimum\\energy\\ \cite{stawicki2016driving}}&	\shortstack{developed four-class\\graphical user\\interface \cite{stawicki2016driving}}& \shortstack{L,R,S\&B for\\turn left\&right,\\ forward and stop}& 93.03\%&-& OM\&SAM
			\\ 				
			\hline
			\shortstack{Teng et al. \cite{teng2017eeg}\\2017}& EEG & \shortstack{denoised EEG\\signals}& \shortstack{CSP, PSD for\\different frequency\\ bands}& RLDA& \shortstack{S\&B for EBC}& 94.00\%& 420 ms& \shortstack{OM,PO\&\\RVM}
			\\ 
			\hline	
			
			\end{tabular}
		\end{turn}
    		%\begin{tablenotes}
    		\small\\\tiny{
    			L= Left, R= Right, S= Steering, B= Brake, RFM= Right Foot Movement, SIM= Self-Initiated Movement, ES= External Stimulus, RTM= Real-time Mode, OM= Offline Mode, POM= Pseudo-Online Mode, RT= Response Time, SAM= Semi-Autonomous Mode, RVM= Real-Vehicle Mode.}
            %\end{tablenotes}
    	\end{table*}
	\newpage

		\begin{table*}
  	    \centering
	    \caption{The second part of reviewed studies in the past 10 years}\label{tabel2}
	    \begin{turn} {90}
		\begin{tabular}{|c|c|c|c|c|c|c|c|c|}
			\hline 
			\shortstack{Authors\\Year} & Signal & Patterns & Features & Classifiers & \shortstack{Task and control\\commands} & Average Accuracy & RT & Modes

			\\ 
			\hline
			\shortstack{Wang et al. \cite{wang2017eeg}\\2017}& EEG & \shortstack{denoised EEG\\signals}&\shortstack{falling-off\\discrimination\\model}& RLDA & \shortstack{S\&B for EBC} & 99.63\% &-&OM
		    \\ 
			\hline
%			\shortstack{Bi et al. \cite{bi2018novel}\\2018}& EEG &P300\&SSVEP& ICA\&CSP& RLDA& \shortstack{S\&B for EBC}& 94.89\% & 540ms & \shortstack{OM, POM\&\\RVM}
 %           \\ 
%			\hline 
			\shortstack{Hernandez et al. \cite{hernandez2018eeg}\\2018}&EEG\&EMG& denoised EEG& time domain& SVM \& CNN& \shortstack{Left Leg movement for\\EBC and OAC} & \shortstack{SVM = 71.10\%,\\CNN = 71.80\%} & 718$\pm${162}ms& OM\& RTM
			 \\
			\hline
			 \shortstack{Bi et al. \cite{bi2018novel}\\2018}& \shortstack{EEG\&external\\sensors}& \shortstack{denoised\\EEG signals}& CSP& RLDA&\shortstack{S\&B for EBC}& 94.89\%& 540ms&\shortstack{OM, PO\& \\RVM}
			 \\
			\hline
	        \shortstack{Yang et al. \cite{yang2018driving}\\2018}& EEG & denoised EEG& FFT\&ICA & \shortstack{long-transformed\\PSD different\\bands\& power\\amplitude} & \shortstack{SVM, KNN and K-\\means with adaptive\\synthetic sample} & \shortstack{S\&B for aggressive\\and unaggressive\\behavior}& 69.50\% & OM
			\\
			\hline
			\shortstack{Lu et al. \cite{lu2018eeg}\\2018}& EEG &SSVEP & PSD& \shortstack{SVM with\\RBF kernel}& \shortstack{S\&B for lane changing,\\path-selection, and\\car-following} & \shortstack{no average, accuracy\\variation\\49.70\% to 100\%} & - & OM\& RTM
			\\ 
			\hline
			\shortstack{Zhuang et al. \cite{zhuang2019ensemble}\\2019} &EEG& ERD\textbackslash ERS& CSP& \shortstack{SVM,CNN\&\\ ensemble model}& \shortstack{L,R,S,B for\\OAV}&	\shortstack{CNN = 87.17\%,\\SVM = 83.25\%,\\ensemble model= 91.75\%}&-& OM\& RTM
			\\ 
			\hline	
			\shortstack{Hekmatmanesh\\et al. \cite{hekmatmanesh2019combination,hekmatmanesh2019optimized} 2019} & EEG& ERD\textbackslash ERS & \shortstack{Wavelet-DFA,long\\-term correlation} & \shortstack{SMSVM with\\GRBF kernel}& \shortstack{S\& B for moving\\forward and brake}& 85.33\% &-& OM
			\\ 
			\hline	
			\shortstack{Hekmatmanesh\\et al. \cite{hekmatmanesh2019eeg} 2019} &EEG& ERD\textbackslash ERS& \shortstack{optimized LLE\\using CTWO}& \shortstack{SSVM with the\\GRBF kernel}& \shortstack{S\&B for moving\\forward and brake}& 68.25\%&-& OM
			\\ 
			\hline	
			\shortstack{Lu et al. \cite{lu2019combined}\\2019}& EEG &SSVEP& \shortstack{optimized\\CSP, PSD}&\shortstack{SVM with RBF\\kernel(one\\against others)}& \shortstack{L, R, S, B for\\ changing lane,\\selecting path\\and following cars}& \shortstack{average accuracy not\\mentioned, accuracy\\ variation 44.20\% to 100\%} &-& OM, RTM
			\\ 
			\hline	
			\shortstack{Lu et al. \cite{lu2019model}\\2019}& EEG& SSVEP& frequency-domain& SVM with MPC& \shortstack{R,L,S for road-\\keeping test and\\obstacle avoidance}& \shortstack{no average values,\\variation from 85.95\%\\to\\99.95\% task success\\<50ms\%-100ms\%} &-& OM, RTM
			\\ 
			\hline	
	    	\shortstack{Nguyen et al. \cite{nguyen2019detection}\\2019 }& EEG& preprocessed EEG&\shortstack{Five PSD\\frequency bands}& NN & \shortstack{S,B, for EBC} & 91.00\% & 600ms& OM, RTM
	    	\\
	    	\hline	
    		\shortstack{Zao et al. \cite{zhao2019identification}\\2019}& \shortstack{signals of brake\\pedal stage} &-& \shortstack{power\& \\autoregressive} & \shortstack{hybrid model\\of GHMM and\\GGAP-RBFNN} & \shortstack{slight, normal\\and EBC} & \shortstack{slight braking = 95.57\%,\\normal braking= 94.69\%,\\emergency brakes = 100ms} &-&\shortstack{OM, RTM\&\\RVM}
    		\\ 
    		\hline	
    		\shortstack{Hekmatmanesh\\et al. \cite{hekmatmanesh2020combination} 2020}& EEG& ERD\textbackslash ERS & CSP,Variance& \shortstack{SSVM with\\GRBF kernel}& \shortstack{S, B for moving\\forward and brake}& \shortstack{OM= 92.70\%,\\RTM = 83.21\%} & 1.5s & OM, RTM
    		\\ 
    		\hline
    			\end{tabular}
    		\end{turn}
    		%\begin{tablenotes}
    		\small\\
    			\tiny{L= Left, R= Right, S= Steering, B= Brake, RFM= Right Foot Movement, SIM= Self-Initiated Movement, ES= External Stimulus, RTM= Real-time Mode, OM= Offline Mode, POM= Pseudo-Online Mode, RT= Response Time, SAM= Semi-Autonomous Mode, RVM= Real-Vehicle Mode.}
        %\end{tablenotes}
    	\end{table*}

    		\newpage
	\begin{table*}
  	    \centering
	    \caption{The first part of BCAV review studies in the past 10 years}\label{tabel3}
	    \begin{turn} {90}
		\begin{tabular}{|c|c|c|c|c|c|c|c|c|}
			\hline 
			\shortstack{Authors\\Year} & Signal & Patterns & Features & Classifiers & \shortstack{Task and control\\commands} &  \shortstack{Average\\Accuracy} & RT & Modes
		    \\ 
			\hline
			\shortstack{Royer et al. \cite{royer2010eeg}\\2010}& EEG & ERD\textbackslash ERS & \shortstack{Cross correlation,\\autoregressive\\spectral amplitude} & Linear classifier & \shortstack{RH, LH, BH,\\rest for\\L,R,U,D\\movements}&-& 2.1s& \shortstack{OM, RTM,\\ RADM}
            \\ 
			\hline 
			\shortstack{Akce et al. \cite{akce2010remote}\\2010}& \shortstack{EEG with\\feedback camera}& ERD\textbackslash ERS& CSAP & \shortstack{HMM with\\belief propagation}& \shortstack{Imaginary RH, LH\\for specifying\\a path for a\\fixed Wings} &-&-&\shortstack{OM, RTM} 
			 \\
			\hline
			\shortstack{Doud et al. \cite{doud2011continuous}\\2011}& \shortstack{EEG with \\camera feedback} & ERD\textbackslash ERS & \shortstack{time-frequency\\PSD} & \shortstack{Amplitude\\threshold of the\\mu band}& \shortstack{RH,LH,BH,rest,\\ thong, feet for\\TR, TL, F,BW,\\U, D}& 85\% &- & \shortstack{OM-RLT,\\RADM}
			 \\
			\hline
			\shortstack{Kim et al. \cite{kim2014quadcopter}\\2014}& \shortstack{EEG\\eye camera} & EEG, eye features & \shortstack{EEG= power\\Eye = contour points\\between the pupil\\and iris}&\shortstack{linear kernel-\\based SVM} & \shortstack{9 points eye\\gazing for 8\\ directions U,D,L,\\R, FW, BW, TR, TL}& \shortstack{OM = 91.67\% \\RTM = 80\%}& \shortstack{eye tracker\\RT = 117ms,\\EEG RT = 84ms}& OM,RTM
			\\
			\hline
			\shortstack{Shi et al. \cite{shi2015brain}\\2015}& \shortstack{eye tracker\\EEG, camera} &ERD\textbackslash ERS &\shortstack{improved cross-\\correlation\\mean, skewness\\kurtosis, minimum,\\maximum, standard\\deviation} &LR & \shortstack{L, R and Idle for\\ L, R, F and camera\\for obstacle avoidance} & - & 94.36\% &OM,SAM
			\\ 
			\hline
			\shortstack{Coenen et al. \cite{coenen2015uav}\\2015}& EEG, EOG& \shortstack{auditory imagination\\and spatial\\navigation\\mental task} & \shortstack{frequency feature\\and regression}& LR & \shortstack{hand joystick or\\foot button box\\movement for\\
            horizontal\&\\Vertical movement}& 85.66\%&- &OM,RTM
			\\ 
			\hline	
			\shortstack{kosmyna et al. \cite{kosmyna2015towards,kosmyna2015adding}\\2015}& EEG\&EMG& \shortstack{ERD\textbackslash ERS\&\\facial patterns} & FastICA, average &\shortstack{adaptive recurrent\\NN, KNN}& \shortstack{L, R, L foot for\\TL, TR, D}&about 75\% &-& OM, RTM
			\\ 
			\hline	
			\shortstack{Khan et al. \cite{khan2015hybrid}\\2015}& EEG, NIRS &\shortstack{ERD\textbackslash ERS\\SSVEP}& \shortstack{PSD of EEG and\\HbO and HbR\\for NIRS} &LDA& \shortstack{6Hz visual\\flickering of\\light and MI\\for U\&D and\\Foreward} &78.33\% &-& OM
			\\
			\hline	
			\shortstack{Lin et al. \cite{lin2015implementing}\\2015}& EEG, EMG& \shortstack{EEG, EMG patterns\\based on facial\\gestures}& \shortstack{preprocessed EEG,\\ EMG} & PCA & \shortstack{small, frown, right\\ and left wink\\ facial gestures\\for R,L,F,\\BW,U,D}&-&-&OM
			\\ 
			\hline	
			\shortstack{Zhang et al. \cite{jacksonquadcopter}\\2018}& \shortstack{EEG, Google glass\\14 external\\sensors} &SSVEP& \shortstack{Fourier Transform\\coefficients}&SVM with RBF& \shortstack{L, R, (both)Foot,\\both hand movement,\\touch pad gesture,\\head posture for\\ R,L,F,BW,U,D}& 97.10\%&	- &OM, RTM
			\\ 
			\hline	
    			\end{tabular}
    		\end{turn}
    		%\begin{tablenotes}
    		\small\\
                \tiny{U = Up, D = Down, F = Forward, BW = Backeward, R = Rotation, L = Left, R = Right, C = clockwise, UC = Un-Clockwise, H = Horizental, V= vertical, HbO = oxygenated hemoglobin, HbR = deoxygenated hemoglobin, RADM = real-aerial drone mode, CSAP = common spatial analytic pattern, MNV = Minimum negative value.}
        %\end{tablenotes}
    	\end{table*}

    \begin{table*}
  	    \centering
	    \caption{The second part of BCAV review studies in the past 10 years}\label{tabel4}
	    \begin{turn} {90}
		\begin{tabular}{|c|c|c|c|c|c|c|c|c|}
			\hline 
			\shortstack{Authors\\Year} & Signal & Patterns & Features & Classifiers & \shortstack{Task and control\\commands} & \shortstack{Average\\Accuracy} & RT & Modes
		    \\ 
			\hline
			\shortstack{Khan et al. \cite{khan2016hybrid}\\2016}& \shortstack{EEG, facial \\ gestures, fNIRS,\\camera} & {SSVEP, EOG}&\shortstack{mean of HbO\\changes, PSD,\\peak and skewness}&LDA& \shortstack{L, R\&eye movement,\\6Hz pattern for\\ U,D, rotation\\stop(obstacle avoidance}&\shortstack{NIRS = 86.66\%\\MI = 87.13\%\\EOG = 86.96\%\\SSVEP = 87.20\%} &- &OM
	    	\\
	    	\hline	
			\shortstack{Khan et al. \cite{khan2017hybrid}\\ 2017}& EEG, fNIRS& \shortstack{patterns related to\\Mental arithmetic,\\mental counting,\\word formation,\\and mental rotation} &  \shortstack{EEG= number of\\peaks, skewness\\fNIRS = min, mean,\\HbO changes\&skewness}&LDA&	U,D,F,R &84.00\% &-&OM, RTM
    		\\ 
    		\hline	
			\shortstack{Khan et al. \cite{khan2017hybrid}\\ 2017}& EEG, fNIRS& \shortstack{patterns related to\\ Mental arithmetic,\\mental counting,\\word formation,\&\\mental rotation\\in frontal, parietal,\& \\visual locations}& \shortstack{EEG= number of\\peaks, skewness\\eye movements from\\EEG= mean,\\number of eye\\blinks, fNIRS=\\min, mean,\\HbO changes}&LDA& \shortstack{L,R, eye movement\\from EEG, IM\\and fNIRS for\\U,D, F,BW,\\TR, TL, CR, UCR}&\shortstack{fNIRS = 75.60\%\\EEG = 86.00\%}&-& \shortstack{OM, RTM,\\RADM}
    		\\ 
    		\hline
    		\shortstack{Zafar et al. \cite{zafar2018drone}\\2018}& fNIRS& \shortstack{patterns related\\to mental arithmetic\\ task}& \shortstack{mean, slope, peak,\\changes of HbO\\and HbR, HbT\\and COE} & \shortstack{SVM and VPA with\\ threshold circle}& \shortstack{Mental arithmetic\\for F with constant\\speed}& \shortstack{SVM =86$\pm${5.40}\% \\VPA =  78$\pm${8.30\%}}&2.3s& \shortstack{OM, RTM,\\RADM}
    	   	\\ 
    		\hline	
    		\shortstack{Zafar et al. \cite{zafar2018initial}\\2018}& fNIRS& \shortstack{patterns related\\mental arithmetic\\and  mental\\counting tasks}& \shortstack{mean, slope, peak,\\ changes of HbO\\and HbR, HbT\\ and COE}& \shortstack{SVM and VPA\\with threshold\\circle}& \shortstack{mental arithmetic\\and mental counting\\F, U, R}& 74.00\%& 2.3s & \shortstack{OM, RTM,\\RADM}
    	   	\\ 
    		\hline	
    		\shortstack{Kavichai et al. \cite{kavichai2019shared}\\2019}& \shortstack{EEG, fNIRS,\\distance measurement\\sensors\& GPS}& \shortstack{the same as\\Khan et al. \cite{khan2017hybrid}}& \shortstack{SCS (combination\\of \cite{khan2017hybrid} with the\\external sensors\\and GPS)}& \shortstack{the same as\\ Khan et al. \cite{khan2017hybrid}}& \shortstack{the same as\\Khan et al. \cite{khan2017hybrid}\\with GPS\&external\\sensors for OAC}& - & - & OM
    	   	\\ 
    		\hline	
    		\shortstack{kryger et al. \cite {kryger2017flight}\\2017} & EEG & \shortstack{IM of hand\\and wrist}& - & - &	\shortstack{U, D, L, R,\\Pitch and Roll} & - & - & OM, RTM
    		\\ 
    		\hline	
    		\shortstack{Wang et al. \cite{wang2018wearable}\\2018}& EEG, HMD & SSVEP & CCA & threshold&\shortstack{flickering at\\11.25 Hz, 6.4Hz,\\7.5 Hz and 9 Hz,\\for U, D, F\\ and TR} &78.00\%& -&OM, RTM
    	   	\\ 
    		\hline	
    		\end{tabular}
    		\end{turn}
    		%\begin{tablenotes}
    		\small\\
                \tiny{U = Up, D=Down, F= Forward, BW=Backeward, R =Rotation, L= Left, R =Right, C= clockwise, UC = Un-Clockwise, H=Horizental, V= vertical, HbO = oxygenated hemoglobin,  HbR = deoxygenated hemoglobin, RADM = real-aerial drone mode, MNV = Minimum negative value.}
        %\end{tablenotes}
       	\end{table*}

\end{document}